\documentclass[twocolumn,twoside,slac_two]{revtex4}
\usepackage{graphicx}
\usepackage{fancyhdr}
\newcommand{\met}{$E_{\rm{T}}\hspace{-1.1em}/$\hspace{0.7em}}
\newcommand{\Znu}{$Z \to \nu\nu$}
\newcommand{\Zmu}{$Z \to \mu\mu$}
\newcommand{\ttb}{$t\bar{t}$~}
\newcommand{\et}{$E_{T}$}
\newcommand{\pt}{$p_{T}$}

\begin{document}

\title{Search for supersymmetry at CMS in the all-hadronic final state}

%

\author{G. Lungu}
\affiliation{The Rockefeller University, New York, USA}

\begin{abstract}
We present a search for supersymmetry (SUSY) in the fully hadronic final state with the CMS detector at the LHC. 
This final state contains at least two jets and a significant transverse energy imbalance due to the neutralinos escaping detection. 
The background to the all-hadronic signature arise from QCD multi-jet production, ttbar and electroweak boson+jet production. 
These backgrounds can be estimated by utilizing a data-driven approach thus enabling a possible discovery of SUSY in the early physics data. 
In addition to the generic search, we also present an analysis which utilizes a new quantity, $\alpha_T$, constructed exclusively from the transverse energies of the jets to effectively eliminate the QCD background.
\end{abstract}

\maketitle

\thispagestyle{fancy}


\section{\label{sec:intro}Introduction}
The supersymmetry~\cite{susy}(SUSY) is a potential symmetry between the fermions and the bosons of the standard model (SM) of particles proposing that each fermion has a bosonic superpartner and each boson has a fermionic superpartner. 
The main attraction of the SUSY models is that they solve the hierarchy and the fine-tuning problems of SM. 
Given that no sparticles have been observed it must be that SUSY is broken such that the sparticles are heavier than the particles. 

One of the most studied SUSY models is the minimally supersymmetric standard model (MSSM) of particles. 
This model is an extension of SM postulating the conservation of the R-parity.
This conservation law results in the pair prodcution of the sparticles and their decay into an odd number of sparticles. 
This means further that the lightest supersymmetric particle (LSP) is stable and interacts very weakly with the matter making it a dark matter candidate.

Another postulate of MSSM is the soft SUSY breaking. 
There are several theoretical approaches to this, but this report will mention only a gravity mediated SUSY breaking (mSUGRA) case.

The strategy used in every search for SUSY is geared towards selecting events with very energetic observables: leptons, photons, jets, missing transverse energy (\met) due to neutrinos and LSP. 
The event selection and the methods used in these searches~\cite{Call} are optimized on the simulation of a set of benchmark points defined by certain values of the SUSY parameters. 
These points are chosen such that a good coverage of the phase space is obtained beyond the current experimental and theoretical limits. 

\section{\label{sec:det}The CMS detector and event clean-up}
The SUSY searches described in this report are using simulated 10~TeV proton-proton collisions followed by the simulation of the detector response to the resulting particles. 
The detector used in these studies is the general purpose CMS~\cite{cms} detector installed at LHC. 
The simulation of this detector is made using the software package GEANT4~\cite{geant} that takes into account a detailed description of the geometry of the detectors, the materials and the magnetic field. 

In order to perform any kind of analysis with the data obtained from this detector, all the observables (i.e. jets, electrons, muons, photons, \met) have to be well understood. 
Given its discriminating power in any search for SUSY, \met needs special attention. 
Besides potential undetected new particles, the sources for \met are many: cracks in the detector, cosmic rays, beam halo, electronic noise, and even SM processes (e.g. semileptonic decay of hadrons). 
At CMS it has been studied that the effect of the cosmic rays and of the beam halo can be minimized by selecting events with significant average electromagnetic fraction ($EEMF>0.1$) and average charged fraction ($ECHF >0.175$). 
\begin{equation}
EEMF=\frac{\sum_{jets} P_{T}^{jets}EMF^{jets}}{\sum_{jets}P_{T}^{jets}}
\end{equation}
\begin{equation}
ECHF=\frac{\sum_{jets}(\sum_{tracks}P_{T}^{tracks})/P_{T}^{jets}}{N_{jets}}
\end{equation}

\section{\label{sec:cmssusy}Search for SUSY program at CMS}
The strategy employed by the CMS collaboration to search for SUSY follows three main directives: model independent classification of the searches based on the topology of the final state, data-driven estimation of the SM backgrounds, and optimization of the event selection based on a set of benchmark signal points. 

\begin{figure}[!htbp]
\begin{center}
\includegraphics[width=0.5\textwidth]{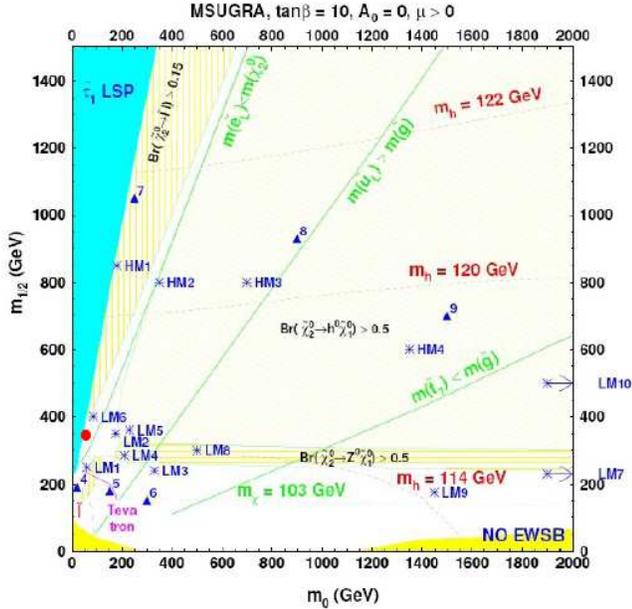}
\caption{The mSUGRA benchmark points used by the CMS collaboration to optimize the selection criteria applied in the SUSY searches.}\label{fig:cmssusy}
\end{center}
\end{figure}

Depending on the physics objects content of the final state, there are eight main SUSY searches at CMS: exclusive n-jet, inclusive 3-jets, photon+\met, single lepton, same-sign dilepton, opposite-sign dilepton, trilepton, and dilepton+photon analysis.
In this report, only the hadronic searches (i.e. exclusive n-jet and inclusive 3-jets) are discussed. 
For each of these analyses the estimation of the various SM backgrounds is performed using techniques that rely as little as possible on the detector simulation. 
Some of these techniques are described briefly in Section~\ref{sec:smbg}. 
In order to enhance the sensitivity of these searches the selection criteria used in these searches are lightly optimized using a set of benchmark SUSY signal points not excluded by theoretical and other hadron colliders experimental constraints. 
The distribution of some of these points is illustrated in Fig.~\ref{fig:cmssusy} for the mSUGRA scenario.
In the following sections the details of the inclusive hadronic search and of the exclusive multi-jet search are discussed.

\subsection{\label{sec:allhad}Inclusive hadronic search}
In this analysis we look for an excess over the SM background in a final state that contains at least three jets and significant \met. 
The trigger used to collect the data contains requirements on the scalar sum of transverse energies (\et) of the jets (HT) and on the magnitude of the vectorial sum of the transverse momentum (\pt) of the jets (MHT). 
\begin{equation}
HT = \sum_{jets} E_{T}^{jets}
\end{equation}
\begin{equation}
\vec{MHT} = \sum_{jets} \vec{p}_{T}^{jets}
\end{equation}
At Level 1 we ask for HT $>$ 200 GeV while at the High Level Trigger (HLT) the HT $>$ 350 GeV and MHT $>$ 100 GeV. 
In addition to the event clean-up described in Section~\ref{sec:det}, the offline analysis requires at least three jets with \et $>$ 30 GeV and pseudo-rapidity ($\eta$) $<$ 3 in absolute value.
The electron or photon contamination of these jets is reduced by requiring the electromagnetic fraction of the jet energy to be $<$ 0.9.

The QCD background can be significantly reduced by selecting a certain region in the $\delta\phi_{1}$-$\delta\phi_{2}$ plane defined such that $R_{1}> 0.5$ and $R_{2}> 0.5$, where
\begin{equation}
R_{1(2)}=\sqrt{\delta\phi_{2(1)}^{2}+(\pi-\delta\phi_{1(2)})^{2}}
\end{equation}
and $\delta\phi_{1(2)}$ is the difference in the polar angles of the leading(subleading) jet and \met.
The contribution from top quark pair production as well as from W boson production is reduced by requiring no isolated tracks in the event.

To reduce the overall SM background, the leading jet is required to have \et $>$ 180 GeV and $|\eta|<$ 1.7, while the second leading jet should have \et $>$ 110 GeV.
For further background reduction we require \met $>$ 200 GeV and $M_{eff}>$ 500 GeV, where $M_{eff}$ is the sum of \met and the \et of the second, third and fourth jets.
This is a prototypical event selection in this analysis that has been optimized using MC samples generated at 14 TeV collisions and described in~\cite{Call}.
For less energetic collisions, the values of the kinematic cuts should be lower than shown above and are currently under study at CMS.

\subsubsection{\label{sec:smbg}SM background}
An important part of the searches for SUSY is the evaluation of the SM background. 
The SM processes have a much larger cross-section than the potential SUSY signal. 
For the LM1 benchmark point ($m_{0}=60$ GeV, $m_{1/2}=250$ GeV, tan$\beta=10$, $A_{0}=0$, $\mu>0$) the leading order (LO) cross-section at 14~TeV is about 42~pb. 
The corresponding LO cross-sections for the most important SM backgrounds are: $\approx$5.6E10~pb for QCD events, $\approx$15E3~pb for events with a $Z$-boson produced in association with jets, and $\approx$800~pb for \ttb events. 
In order to minimize the systematic uncertainties due to potential differences between data and simulation, the strategy behind the evaluation of these backgrounds relies on data driven methods. 
Also with these methods there is no need to calibrate the simulation to the data. 

The most important SM background is due to QCD processes. 
Using the cuts on the angle between the leading jets and \met(see Section~\ref{sec:allhad}), about 80$\%$ of the QCD events can be rejected, while keeping about 90$\%$ of the SUSY events. 
The remaining QCD content can be evaluated directly from data via two proposed methods: smearing method and ABCD method. 

The smearing method, developed at ATLAS~\cite{Asm},  relies on the parameterization of a response function from multijet events with high \met values and \met aligned with one of the jets. 
The response function is defined for each jet as 
\begin{equation}
R=\frac{1-p_{T}^{jet}cos(jet,E_{\rm{T}}\hspace{-1.1em}/\hspace{0.7em})}{|p_{T}^{jet}+E_{\rm{T}}\hspace{-1.1em}/\hspace{0.7em}|^{2}}
\end{equation}
and it used to smear the jets from events with low \met. 
The smearing of the jets will result in artificially created \met used to estimate the real \met distribution. 
The normalization is obtained from the multijet data events with low \met.

The ABCD method, pursued at CMS, uses two uncorrelated observables (\met variable and the minimum azimuthal angle between its direction and that of the three leading jets).
The plane formed by these two variables is split in four regions: A, B, C and D, such that the SUSY signal is contained in region C. 
Given that the two observables are uncorrelated, the number of QCD events in region C can be derived from the number of QCD events in the other regions: $C=DxB/A$. 
For this method to work it is important to have little SUSY content in region A, B and D. 

\begin{figure}[!htbp]
\begin{center}
\includegraphics[width=0.5\textwidth]{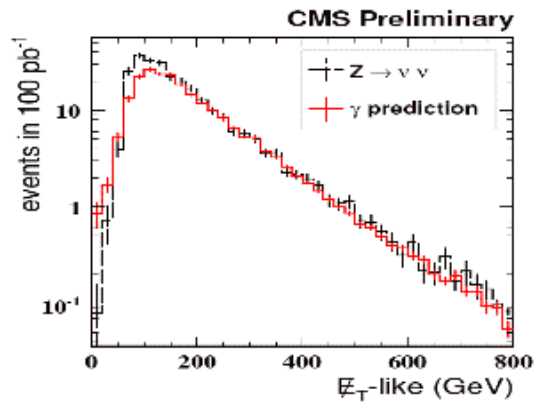}
\caption{The comparison between the \met distribution from \Znu~+jets MC sample and the corresponding distributions using $\gamma$+jets events.}\label{fig:3}
\end{center}
\end{figure}

Another important SM background is produced by \Znu~+jets events. 
This background is irreducible and there are few proposed ways to measure it. 
One method relies on the measurement of \met from \Zmu~+jets events where the muons are ignored. 
The normalization is set by taking into account the theoretical ratio of cross-sections between the \Znu~and \Zmu~processes. 
Another method, developed at CMS~\cite{znuphot}, relies on $\gamma$+jets events with the photon being ignored in the event. 
This method benefits from a larger data sample, while the disadvantage is due to the uncertainty on the normalization. 
The plot in Figure~\ref{fig:3} shows the comparison between the \met distribution from \Znu~+jets MC sample and the corresponding distributions using $\gamma$+jets events. 
The \Znu~+jets events background can also be estimated from leptonic $W$ decays where the lepton is ignored. 
This method~\cite{znuphot} also benefits from a larger statistics compared to \Zmu~+jets, but less than $\gamma$+jets events. 

\subsubsection{Sensitivity reach}
\begin{figure}[!htbp]
\begin{center}
\includegraphics[width=0.5\textwidth]{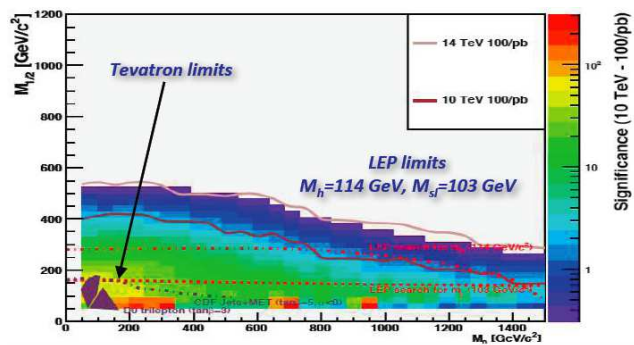}
\caption{Discovery potential in the inclusive 3 jets + \met final state at CMS.}\label{fig:cmssens}
\end{center}
\end{figure}

In Fig.~\ref{fig:cmssens} it is shown the 5-sigma discovery limit in the mSUGRA m$_{0}$-m$_{1/2}$ plane for the final state with at least three jets and \met.
These curves are shown assuming 100~pb$^{-1}$ of good data collected by CMS for 14 TeV collisions (light color) and 10 TeV collisions (dark). 
The limits obtained from the Tevatron experiments (solid, dark) and from LEP (dashed lines) are also shown.
The curve for the 10 TeV collisions has been obtained using the same event selection used for the 14 TeV case. 
We conclude that for this final state there is a very good chance of discovering SUSY with several tens of pb$^{-1}$ of understood data.

\subsection{Exclusive two jets channel}
\begin{figure}[!htbp]
\begin{center}
\includegraphics[width=0.5\textwidth]{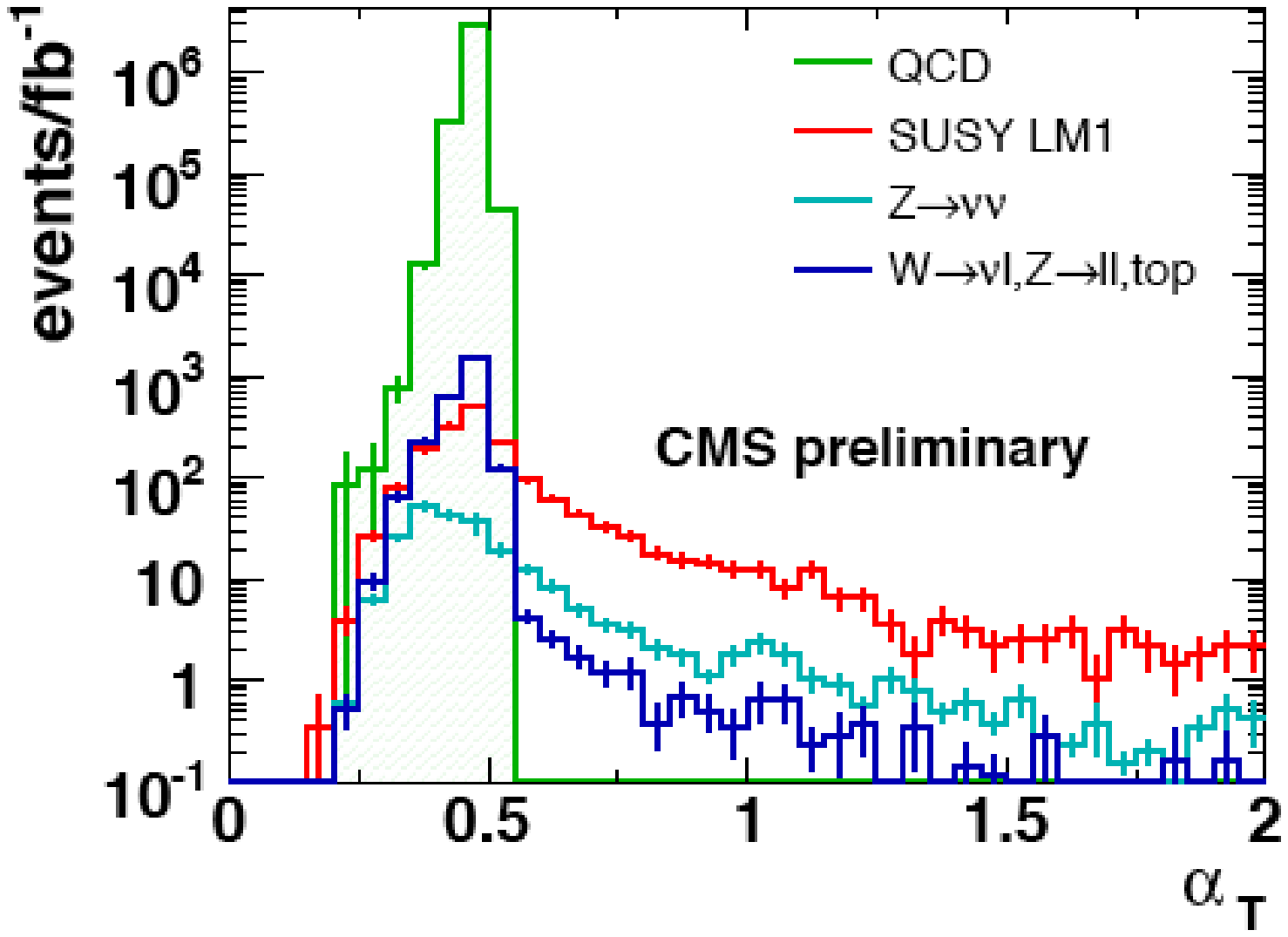}
\caption{Distribution of the $\alpha_{T}$ variable calculated in events with  twojets.}\label{fig:alphaT}
\end{center}
\end{figure}

It expected that in the early days the \met observable will be poorly understood. 
As studied  in~\cite{djet}, a new discriminating variable ($\alpha_{T}$) is used instead of \met: the ratio between the transverse energy of the second leading jet and the transverse invariant mass of the two leading jets. 
The search is performed in events with the two leading jets having \pt $>$ 50 GeV/c, and no hard third jet (\pt $<$ 50 GeV/c) or hard leptons (\pt $<$ 10 GeV/c). 
The dominant QCD dijet background can be dramatically reduced by requiring $\alpha_{T}>0.55$ as it can be seen in Fig.~\ref{fig:alphaT}.
An additional cut on the HT $>$ 500 GeV keeps the SM backgrounds under control.

The remaining backgrounds as listed in Fig.~\ref{fig:alphaT} are estimated with a data-driven method using an ABCD-like technique where the variables of interest are $\alpha_{T}$ and $|\eta|$ of the leading jet. 
As it can be seen in Fig.~/ref{fig:alphaTeta}, the signal (LM1) is expected to have the leading jet mostly in the central region ($|\eta|<$ 2.5), while the SM background extends more forward in pseudo-rapidity.
 
\begin{figure}[!htbp]
\begin{center}
\includegraphics[width=0.5\textwidth]{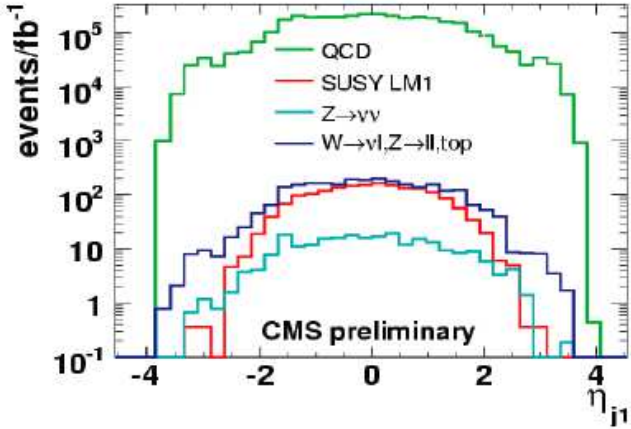}
\caption{Distribution of the pseudo-rapidity of the leading jet for SUSY signal (LM1) and various SM backgrounds.}\label{fig:alphaTeta}
\end{center}
\end{figure}

\begin{figure}[!htbp]
\begin{center}
\includegraphics[width=0.5\textwidth]{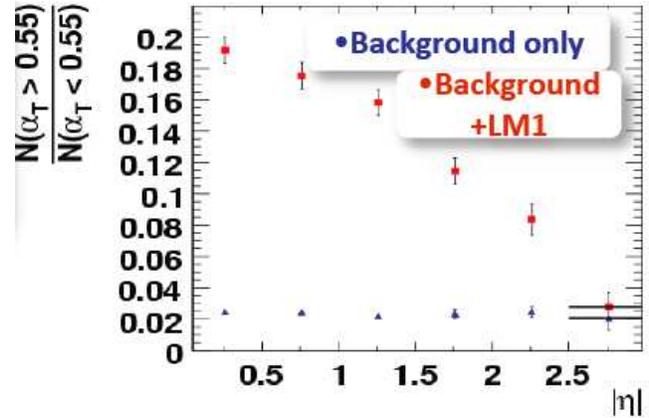}
\caption{The ratio of the number of events with $\alpha_{T}>0.55$ to the number of events with $\alpha_{T}<0.55$ as a function of the pseudo-rapidity of the leading jet. The triangles correspond to the SM background events while the squares represent the mix of signal (LM1) and the various SM backgrounds.}\label{fig:alphaTbg}
\end{center}
\end{figure}

The signal region (C) is defined by $\alpha_{T}>0.55$ and $|\eta_{1}|<$2.5, while the control regions are defined as follows: region A has $\alpha_{T}<0.55$ and $|\eta_{1}|>$2.5, region B has $\alpha_{T}>0.55$ and $|\eta_{1}|>$2.5, and region D has $\alpha_{T}<0.55$ and $|\eta_{1}|<$2.5
The amount of background in the signal region is then equal to $Dx B/A$.
Fig.~\ref{fig:alphaTbg} shows that for background (triangles) the fraction $B/A$ is constant as a function of the pseudo-rapidity of the leading jet, while adding signal (squares) introduces a dependence. 
It has to be noted here that relaxing the HT cut results in a larger signal contamination in regions A and B which will create larger systematic uncertainties on the background estimate.

With this background estimation method, the signal to background ratio is expected to be of order 6 and SUSY signal to be $\approx$44 events for 100~pb$^{-1}$ of data.

\subsubsection{Generalization to n-jets}
\begin{figure}[!htbp]
\begin{center}
\includegraphics[width=0.5\textwidth]{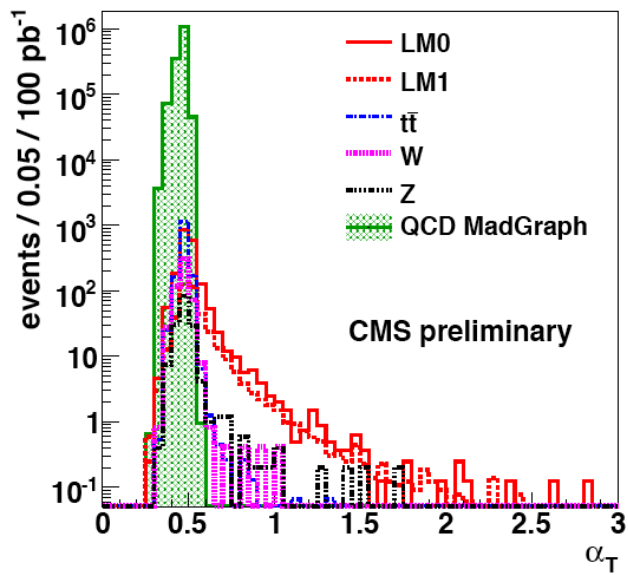}
\caption{Distribution of the $\alpha_{T}$ variable calculated in events with n jets.}\label{fig:alphaTn}
\end{center}
\end{figure}

The techniques used in the search performed on the sample of events with exactly two jets can be applied on the sample with more jets by generalizing the definition of $\alpha_{T}$.
This can be done by combining the jets into two groups such that the difference ($\Delta H_{T}^{min}$) between the HT calculated in each group is minimized. 
The generalized $\alpha_{T}$ is defined as $(H_{T}-\Delta H_{T}^{min})/2M_{T}$, where HT is computed using all the jets used in the analysis and $M_{T}$ is their transverse invariant mass. 
In Fig.~\ref{fig:alphaTn} it is shown that the dominant QCD background can be effectively reduced by cutting on this generalized variable. 

As detailed in~\cite{njet}, this analysis has been performed in sample of events with 3, 4, 5, and 6 jets in the final state. 
The trigger used to select these events requires at L1 that the leading jet has \et $>$ 70 GeV and at HLT \et $>$ 110 GeV. 
The jets used in this analysis are required to have \et $>$ 50 GeV and $|\eta|<$3, while the leading two jets must have \et $>$ 100 GeV and the leading jet $|\eta|<$2. 
In order to eliminate leptonic events the muons and electrons in these events should have \pt $<$ 10 GeV/c.

The remaining backgrounds are further reduced by requiring the events to have HT $>$ 350 GeV. 
The contribution from the QCD background is diminished by requiring that $\alpha_{T}>$ 0.55.  
To reduce the QCD contribution to large $\alpha_{T}>$ values, it is required that the ratio of MHT calculated using jets with \et $>$ 50 GeV to the MHT using jets with \et $>$ 30 GeV is less than 1.25. 

As in the case of the exclusive two jets analysis, the signal to background ratio is expected to be of order 6 and SUSY signal to be $\approx$52 events for 100~pb$^{-1}$ of data.

\subsection{Conclusion} 
The CMS collaboration has a search for SUSY program covering all possible final states and it is using a variety of tools and methods to this end. 
In particular, the searches for SUSY in the fully hadronic channels are well advanced and the most sensitive. 
This sensitivity is in part due to the data-driven approaches to estimate the SM backgrounds thus minimizing the systematic uncertainties. 
The studies described in this report indicate that SUSY can be discovered in the fully hadronic final state with less than 100~pb$^{-1}$ of understood data. 


\bigskip 

\end{document}